\def\Statusstring{  }

\documentclass[12pt,letter]{article}

\usepackage{amsmath, amsthm, amssymb}
\usepackage{amssymb,epsfig}

\usepackage{hyperref}
\hypersetup{bookmarks=true}

\makeatletter

\gdef\@punct{.\ \ }  % Punctuation after run-in section heading
\def\@sect#1#2#3#4#5#6[#7]#8{%
  \ifnum #2>\c@secnumdepth
     \def\@svsec{}
  \else
     \refstepcounter{#1}\edef\@svsec{%
     \ifnum #2>0{{\csname the#1\endcsname}}.\fi%
    \hskip .5em}
  \fi
  \@tempskipa #5\relax
  \ifdim \@tempskipa>\z@
     \begingroup #6\relax
       \@hangfrom{\hskip #3\relax\@svsec}{\interlinepenalty \@M #8\par}
     \endgroup
     \csname #1mark\endcsname{#7}
     \addcontentsline{toc}{#1}{\ifnum #2>\c@secnumdepth\else
          \protect\numberline{\csname the#1\endcsname}\fi#7}
  \else
     \def\@svsechd{#6\hskip #3\@svsec #8\@punct\csname
#1mark\endcsname{#7}
     \addcontentsline{toc}{#1}{\ifnum #2>\c@secnumdepth \else
          \protect\numberline{\csname the#1\endcsname}\fi#7}}
  \fi
  \@xsect{#5}}

\def\@ssect#1#2#3#4#5{\@tempskipa #3\relax
  \ifdim \@tempskipa>\z@
     \begingroup #4\@hangfrom{\hskip #1}{\interlinepenalty \@M
#5\par}\endgroup
  \else \def\@svsechd{#4\hskip #1\relax #5\@punct}\fi
  \@xsect{#3}}

\newcommand{\imp}{\mathop{\mathrm{imp}}\nolimits}

\def\qed{\hskip 3pt \hbox{\vrule width4pt depth2pt height6pt}}

\newtheorem{Lemma}{Lemma}
\newtheorem{Example}[Lemma]{Example}
\newtheorem{Theorem}[Lemma]{Theorem}
\newtheorem{Proposition}[Lemma]{Proposition}

\newtheorem{Construction}[Lemma]{Construction}
\newtheorem{Algorithm}[Lemma]{Algorithm}
\newtheorem{Remark}[Lemma]{Remark}

\usepackage{calc}
\usepackage{tikz}
\usetikzlibrary{decorations.markings}
\tikzstyle{vertex}=[circle, draw, inner sep=0pt, minimum size=6pt]
\newcommand{\vertex}{\node[vertex]}
\tikzset{->-/.style={decoration={
markings,
mark=at position #1 with {\arrow{>}}},postaction={decorate}}}

\begin{document}

\title{Performance Guarantees of Distributed Algorithms for QoS in Wireless Ad Hoc Networks}

\author{Ashwin~Ganesan%
  \thanks{The author is an Independent Researcher (ashwin.ganesan@gmail.com).  The author was with the School of Computer Science and Engineering, Vellore Institute of 
Technology, Vellore 632014, Tamilnadu, India.} 
%53 Deonar House, Deonar Village Road, Mumbai-88, India; 
% A.G.~is the corresponding author.}
}

\date{}

\maketitle

\vspace{-8.0cm}
\begin{flushright}
  \texttt{\Statusstring}\\[1cm]
\end{flushright}
\vspace{+5.0cm}

\begin{abstract}
\noindent  
Consider a wireless network where each communication link has a minimum bandwidth quality-of-service requirement.  Certain pairs of wireless links interfere with each other due to being in the same vicinity, and this interference is modeled by a conflict graph.  
Given the conflict graph and link bandwidth requirements, the objective is to determine, using only localized information, whether the demands of all the links can be satisfied.  At one extreme, each node knows the demands of only its neighbors; at the other extreme, there exists an optimal, centralized scheduler that has global information.  The present work interpolates between these two extremes by quantifying the tradeoff between the degree of decentralization and the performance of the distributed algorithm.  This open problem is resolved for the primary interference model, and the following general result is obtained: if each node knows the demands of all links in a ball of radius $d$ centered at the node, then there is a distributed algorithm whose performance is away from that of an optimal, centralized algorithm by a factor of at most $(2d+3)/(2d+2)$. The tradeoff between performance and complexity of the distributed algorithm is also analyzed.  It is shown that for line networks under the protocol interference model, the row constraints are a factor of at most $3$ away from optimal. Both bounds are best possible. 

\end{abstract}

\bigskip
\noindent\textbf{Index terms} --- graph theory, wireless ad hoc networks, distributed algorithms, admission control, quality-of-service, imperfection ratio, row constraints, primary interference model, protocol interference model, line networks

%\newpage
%---------------------------------------------------------------
%\vskip 0.3in
% \newpage
\tableofcontents

%\vspace{+0.5cm}

\section{Introduction}

Inelastic applications have fixed requirements for network resources which are not adaptable to network performance. Examples include real-time applications such as voice and video, where the transmitted information must be received without much delay.  The best-effort model of the internet, which does not make guarantees on timely delivery, is not sufficient for such applications.  When an inelastic application makes a demand for network resources, the admission control problem is to decide whether to admit the flow.  If the new demand can be satisfied without disrupting the service already promised to existing flows, then the new flow is admitted.  Otherwise, the network signals it is busy, and the application attempts again at a later time.  The present work considers inelastic applications whose quality-of-service (QoS) requirements are specified in terms of the minimum bandwidth required by communication links in the network.  The links are wireless, and wireless links in the same vicinity contend for the shared wireless medium.  The bandwidth requirements are for links between nodes which are within communication radius of each other; the flows in this work refer to single-hop flows.  

Consider an ad hoc network where nodes in the same vicinity directly communicate with each other, without any centralized infrastructure.  The network can be modeled by an undirected graph $G=(V,L)$, where $V$ is a set of nodes and $L$ is a set of wireless links.  The interference is modeled by a conflict graph $G_c = (L, L')$ whose vertices are the wireless links, and two wireless links are adjacent vertices in the conflict graph if and only if they cannot be simultaneously active due to interference.  Suppose that each link $\ell$ makes a demand for bandwidth $f(\ell)$ b/s and that the total available bandwidth of the shared wireless medium is $C$ b/s.  In the admission control problem studied in this paper, the objective is to determine, given the conflict graph $G_c=(L,L')$, link demand vector $(f(\ell): \ell \in L)$ and total bandwidth $C$, whether demands of all the links can be satisfied.  The scheduling problem, which is to obtain a link schedule satisfying the demands, is not considered in the present work.  

Conflict graphs, which were introduced in \cite{Jain:Padhye:etal:03},  can be constructed from constraints imposed by the network's MAC (medium access control) protocol.   For example, in IEEE 802.11 MAC protocol-based networks, if a node $i$ is communicating with node $j$, then all nodes which are neighbors of $i$ or $j$ must remain idle while this communication takes place.  This implies that two wireless links correspond to adjacent vertices in the conflict graph whenever they are at most one hop away from each other in the network connectivity graph. Another type of interference model is the primary interference model. In this model, if $G=(V,L)$ is a network graph, then two wireless links $\ell, \ell'$ interfere with each other if and only if they share an endpoint in common.   The \emph{line graph} of a graph $G=(V,L)$, denoted by $L(G)$, is defined to be the graph with vertex set $L$, and two vertices $\ell_i, \ell_j \in L$ are adjacent in the line graph whenever the corresponding two links $
\ell_i, \ell_j$ of the network graph share a common endpoint in $G$.  Thus, under the primary interference model, the conflict graph $G_c$ is the line graph of the network graph $G$, i.e. $G_c = L(G)$.  A \emph{claw} in a graph is an induced subgraph isomorphic to $K_{1,3}$, the star graph with $3$ leaves. It can be seen that line graphs are claw-free.  Because of the additional structure imposed on the conflict graph by the primary interference model, the general resource allocation problem of assigning colors to the vertices of an arbitrary conflict graph is reduced to the special case of coloring the edges of the network graph.  This problem has been well-studied in the literature \cite{Hajek:1984} \cite{Hajek:Sasaki:1988} and this interference model arises in some practical contexts \cite{Tassiulas:Sarkar:02} \cite{Miller:etal:2000} \cite{Kodialam:Nandagopal:03} \cite{Kodialam:Nandagopal:05}. 

At one extreme, a centralized and optimal solution to the admission control problem exists: if the topology of the entire network and its conflict graph is known to a particular node and the demands of all the wireless links are also known to this center node, then this node can ascertain whether there exists a feasible schedule satisfying all the demands.  However, there is a cost associated with communicating information from distant nodes to a center node, and determining an optimal solution is usually computationally expensive.  Hence, it is desired that the solution be as decentralized as possible and that the admission control algorithm be efficient.  

At the other extreme, each node in the conflict graph knows only the demands of all other nodes (wireless links) interfering with it.  Sufficient conditions for distributed flow admission control in this setting include the row constraints \cite{Gupta:Musacchio:Walrand:07} \cite{Hamdaoui:Ramanathan:05} and the scaled clique constraints \cite{Gupta:Musacchio:Walrand:07} (see Section~\ref{sec:model:problem:formulation} below for their definitions).  Because only localized information is used in these admission control mechanisms, these conditions are conservative in the sense that they overestimate the resources required to satisfy the given demands.  In other words, flows may be denied admission even though they are physically feasible, i.e. feasible as per a centralized scheduler.  Consequently, network resources are potentially underutilized.  Ideally, the acceptance rate for flows of the distributed admission control algorithm is as high as that of an optimal, centralized mechanism.

The present paper interpolates between the above two extremes and quantitatively characterizes the trade-off between the performance of the distributed algorithm and the extent of decentralization for arbitrary networks under the primary interference model.   Also, the worst-case performance of the row constraints is shown to be a bounded factor away from optimal for line networks under the protocol interference model.

%====================================

\subsection{Model and Problem Formulation} \label{sec:model:problem:formulation}

Consider a wireless network represented by a simple, undirected graph $G=(V,L)$, where $V$ is a set of wireless transceivers, also called nodes, and $L$ is a set of wireless links.  The edge set $L$ consists of pairs of nodes which are within communication radius of each other.  Nodes in the same vicinity cannot be simultaneously active due to wireless interference, and this interference is modeled by a conflict graph $G_c=(L, L')$, as follows.  The vertex set of the conflict graph is the set $L$ of communication links in the network.  Two wireless links $\ell_i, \ell_j \in L$ are adjacent vertices in $G_c$ whenever they cannot be simultaneously active due to wireless interference.   In the sequel, $G=(V,L)$ is referred to as the \emph{network graph} and $G_c = (L, L')$ as the \emph{conflict graph}.  Also, the notation and terminology from graph theory used henceforth is standard \cite{Bollobas:1998}; the complete graph on $n$ vertices is denoted by $K_n$, the star graph with $n$ leaves is denoted by $K_{1,n}$, and the cycle graph on $n$ vertices ($n \ge 3$), also called the $n$-cycle graph, is denoted by $C_n$. Thus, the triangle graph is the graph $C_3$ (or $K_3$).

The quality-of-service (QoS) requirements are specified in terms of a minimum bandwidth requirement for each link.  More specifically, suppose each link $\ell \in L$ makes a demand to transmit information at a certain data rate $f(\ell)$ b/s.  Suppose the maximum transmission rate of link $\ell$ is $C(\ell)$ b/s.  Then, the demand for link $\ell$ can be satisfied if link $\ell$ can be active for a fraction $\tau(\ell): = f(\ell) / C(\ell)$ of every unit of time.  It is assumed throughout that $\tau(\ell)$ is a rational number.

The scheduling and flow admission control problems are now formally stated.  An independent set in a graph is a subset of vertices that  are pairwise nonadjacent.  Because nonadjacent vertices in the conflict graph $G_c$ represent pairs of links which do not interfere with each other, an independent set in $G_c$ corresponds to a set of links which can be simultaneously active.  Let $\mathcal{I}(G_c)$ denote the set of all independent sets in $G_c$.  A {\em schedule} is a map $t: \mathcal{I}(G_c) \rightarrow \mathbb{R}_{\ge 0}$ which assigns a time duration $t(I)$ to each independent set $I \in \mathcal{I}(G_c)$.  When schedule $t$ is implemented, all links in an independent set $I$ will be simultaneously active for duration $t(I)$.  The \emph{(total) duration} of the schedule is $T = \sum_{I \in \mathcal{I}(G_c)} t(I)$, and a particular link $\ell \in L$ is active for total duration $\sum_{I \in \mathcal{I}(G_c): \ell \in I} t(I)$.  If the total duration of each link $\ell \in L$ is at least $\tau(\ell)$, then we say there exists a schedule of duration $T$ satisfying link demand vector $(\tau(\ell): \ell \in L)$. An optimal schedule for $\tau$ is a schedule of minimum duration satisfying demand $\tau$. 

Given a conflict graph $G_c = (L, L')$ and link demand vector $\tau = (\tau(\ell): \ell \in L)$, where $\tau(\ell)$ represents the fraction of every unit of time link $\ell$ is required to be active, let $T^*(G_c, \tau)$ denote the minimum duration of a schedule satisfying demand $\tau$. The demand $\tau$ is said to be {\em feasible within duration $T$} if there exists a schedule of duration at most $T$ satisfying demand $\tau$, i.e. if $T^*(G_c, \tau) \le T$.  The demand $\tau$ is said to be {\em feasible} if there exists a schedule of duration at most $1$ satisfying demand $\tau$.  In the distributed admission control problem studied in the present paper, the problem is to determine, given $(G_c, \tau)$, whether there exists a schedule of duration at most $1$ satisfying demand $\tau$, i.e. whether $T^*(G_c, \tau) \le 1$, using only localized information.  The problem of obtaining a schedule which satisfies this demand is not studied in this work.  The independent set polytope $P_I = P_I(G_c)$ is defined to be the convex hull of the characteristic vectors of the independent sets in $G_c$.  Thus, $P_I(G_c)$ is the set of all link demand vectors which are feasible.

The row constraints, scaled clique constraints, and the degree condition are sufficient conditions for admission control which can be implemented in a distributed manner \cite{Gupta:Musacchio:Walrand:07} \cite{Hamdaoui:Ramanathan:05}. These conditions are briefly defined now.  Suppose $G_c = (L, L')$ is a conflict graph and $(\tau(\ell): \ell \in L)$ is a demand vector, where $\tau(\ell)$ is the fraction of each unit of time that link $\ell$ demands to be active.   A sufficient condition for $\tau$ to be feasible is the \emph{row constraints}, which is the condition $\tau(\ell)+\tau(\Gamma(\ell)) \le 1$, for all $\ell \in L$.  Here, $\tau(A)$ denotes $\sum_{a \in A} \tau(a)$, where $A \subseteq L$, and $\Gamma(\ell)$ denotes the set of neighbors in $G_c$ of vertex $\ell$.  Intuitively, this condition is similar to the greedy coloring algorithm for weighted graphs, or can be thought of as a maximal scheduler that assigns to a link any time duration that is not already assigned to other links interfering with it.  Let $d(\ell)$ denote the degree in $G_c$ of vertex $\ell$; thus, $d(\ell)$ is the number of other wireless links that interference with link $\ell$.  A demand vector $\tau$ is feasible if $\tau(\ell)(d(\ell)+1) \le 1$ for all $\ell \in L$; this sufficient condition for admission control is called the degree condition.  

A necessary condition for $\tau$ to be feasible is the clique constraint $T_{\mbox{clique}}(G_c, \tau) \le 1$, where $T_{\mbox{clique}}(G_c, \tau)$ denotes the maximum of $\tau(K)$ over all cliques $K$ in the conflict graph $G_c$.  The \emph{imperfection ratio} of a graph $G_c$, denoted by $\mbox{imp}(G_c)$, is defined to be $\sup_{\tau \ne 0} \{T^*(G_c, \tau) / T_{\mbox{clique}}(G_c, \tau) \}$, where the supremum is over all non-zero integral vectors $\tau$.  The numerator is the exact amount of resources required to satisfy demand $\tau$; the denominator is a lower bound estimate of the resource required to satisfy demand $\tau$, as determined by a particular distributed algorithm - the clique constraints.  Their ratio is the factor by which the estimate is away from optimal.  The maximum possible value of this ratio, namely the imperfection ratio, characterizes the worst-case performance of the distributed algorithm.  A sufficient condition for $\tau$ to be feasible is the \emph{scaled clique constraint} $\mbox{imp}(G_c) T_{\mbox{clique}}(G_c, \tau) \le 1$.  

An equivalent formulation of the admission control problem in terms of the fractional chromatic number of a weighted graph is now given.  For an introduction to fractional graph theory, the reader is referred to \cite{Scheinerman:Ullman:1997} \cite{Godsil:Royle:2001}. Let $G=(V,E)$ be a graph with vertex set $V = \{v_1,\ldots,v_n\}$, and let $\{I_1, \ldots, I_K\}$ denote the set of all independent sets in $G$.  Define the vertex-independent set incidence matrix $B = [b_{ij}]$ of $G$ by $b_{ij} = 1$ if $v_i \in I_j$ and $b_{ij} = 0$ if $v_i \notin I_j$.  Thus, $B$ is a 0,1-matrix of size $n \times K$. The chromatic number of $G$, denoted $\chi(G)$, is defined to be the optimal value of the integer linear program
$$ \min 1^T x \mbox{ subject to } Bx \ge 1, x \in \{0, 1 \}^K. $$
Relaxing the condition that $x$ be integral gives the linear program
$$ \min 1^T x \mbox{ subject to } Bx \ge 1, x \ge 0, $$
whose optimal value is called the {\em fractional chromatic number} of $G$ and is denoted  $\chi_f(G)$. If $\tau = (\tau(v): v \in V)$ is a set of nonnegative weights on the vertex set $V$, then the \emph {fractional chromatic number} of the weighted graph $(G, \tau)$, denoted $\chi_f(G, \tau)$, is defined to be the optimal value of the linear program
$$ \min 1^T x \mbox{ subject to } Bx \ge \tau, x \ge 0. $$
Given a conflict graph $G_c = (L, L')$ and link demand vector $\tau = (\tau(\ell): \ell \in L)$, it is seen that the minimum duration of a schedule satisfying demand $\tau$ is equal to the fractional chromatic number $\chi_f(G_c, \tau)$.   %Thus, the distributed admission control problem is: given $(G_c, \tau)$, determine whether $\chi_f(G_c, \tau) \le 1$ using only localized information.  

The problem of computing the fractional chromatic number of a weighted graph can be shown to be NP-hard \cite{Grotschel:Lovasz:Schrijver:1981} by using the ellipsoid algorithm to give a polynomial transformation between the said problem and the maximum independent set problem. Even though the fractional chromatic number $\chi_f(G_c,\tau)$ can be defined as the value of a linear program and linear programs can be solved in polynomial-time in the size of input, the number of independent sets in the conflict graph can be exponential in the size of the graph.  Hence, the size of the linear program can be exponential in the size of the graph.   A special case of this problem is the case of uniform demands; this problem is equivalent to computing the fractional chromatic number of a graph (an unweighted graph), and this problem is also NP-hard and is known to be hard to even approximate  \cite{Lund:Yannakakis:1994}.  These negative results make it all the more important to design efficient, distributed mechanisms for admission control with provable performance guarantees.

The worst-case performance of a sufficient condition for admission control is defined as follows.  Recall that the independent set polytope $P_I$ of the conflict graph is the set of all link demand vectors $\tau$ which are feasible. A necessary and sufficient condition for $\tau$ to be feasible is that $\tau \in P_I$.  In general, determining whether $\tau \in P_I$ is computationally intensive and requires global information.  Hence, one is interested in obtaining sufficient conditions for admission control that can be implemented efficiently and in a distributed manner.  
Let $S$ be any sufficient condition for admission control (particular examples of sufficient conditions are given in Section~\ref{sec:dist:algo:primary:interference} and Section~\ref{sec:row:line:networks:protocol:interference:model}).  Let $P_S$ denote the set of all link demand vectors $\tau$ which satisfy condition $S$.  Then, $P_S \subseteq P_I$. Because condition $S$ uses only localized information, it is often suboptimal - it is conservative in the sense that it overestimates the amount of resources required to satisfy a given demand $\tau$.  One can scale the resource requirements in the sufficient condition $S$ to obtain a necessary condition. The worst-case performance of the sufficient condition $S$ is defined to be the smallest $\alpha$ such that $P_I \subseteq \alpha P_S$. The sufficient condition $S$ is said to be a factor of at most $\alpha$ away from optimal.

%====================================
\subsection{Summary of Results}

The main contributions of this paper are as follows.

\begin{enumerate}
 \item \emph{A new distributed algorithm for admission control.}  A distance-$d$ distributed algorithm for admission control is given for arbitrary networks when the interference is modeled using primary interference.  The sufficient condition given can be implemented in a distributed fashion in the sense that each node needs to know the demands of only those links which are at most $d$ hops away. The parameter $d$ is the degree of centralization, and one can choose this parameter to be any value between $0$ and the diameter of the network graph.  In the $d=0$ case, a new proof is given which does not rely on Shannon's upper bound on the chromatic index of multigraphs.

 \item \emph{Worst-case performance analysis of the distance-$d$ distributed algorithm.}  The distance-$d$ distributed algorithm is shown to be a factor of at most $(2d+3)/(2d+2)$ away from an optimal, centralized algorithm. This extends previous results in \cite{Shannon:1949} which focuses on the $d=0$ case, and results in \cite{Ganesan:2009} \cite{Ganesan:WN:2014} where the $d=1$ case was studied. This result also resolves the open problem posed in \cite[p. 1333]{Ganesan:WN:2014}, which asks to quantitatively characterize the tradeoff between the degree of localization of the distributed algorithm and the performance of the distributed algorithm. There is also a tradeoff between performance and complexity, which is analyzed.
 
 \item \emph{Performance of row constraints in line networks under protocol interference model.}  It is shown that the row constraints are a factor of at most $3$ away from optimal for line networks under the protocol interference model (cf. Theorem~\ref{thm:line:networks:sigma:le:3}).  In \cite{Kose:Medard:201711} \cite{Kose:etal:201712}, it was claimed that the conflict graphs arising in this context are claw-free. However, this seems to be incorrect - in the present paper, an example is given to show that there exist line networks for which, under the protocol interference model, the conflict graph contains a claw.  Consequently, polynomial time scheduling algorithms that exist for claw-free graphs \cite{Minty:1980} \cite{Nakamura:Tamura:2001} \cite{Faenza:etal:2014} are not applicable. In the present paper, it is also shown that the bound of $3$ is best possible. That is, the bound is tight in the sense that no smaller value can characterize the worst-case performance of the row constraints because there exist line networks for which the row constraints are a factor of exactly $3$ away from optimal; see Theorem~\ref{thm:countereg:KoseMedard} for a proof of the existence of line networks for which the conflict graph contains a claw. 
\end{enumerate}

The rest of this paper is organized as follows. Section~\ref{sec:related:work} mentions some of the literature related to the present work.  In Section~\ref{sec:dist:algo:preliminaries}, the distance-$d$ distributed algorithms for the cases $d=0$ and $d=1$ are studied. In Section~\ref{sec:distance:d:distributed}, a new distance-$d$ distributed algorithm is proposed, and in Section~\ref{sec:distance:d:distr:performance} the algorithm's performance is analyzed. Section~\ref{sec:intuition:dist:implementation} contains a further discussion of the performance and complexity analysis of the distributed algorithm. In Section~\ref{sec:row:line:networks:protocol:interference:model}, the worst-case performance of the row-constraints for line networks under the protocol interference model is analyzed.  Section~\ref{sec:concluding:remarks} contains concluding remarks.

%====================================
\section{Related Work} \label{sec:related:work}

The design of distributed admission control mechanisms and distributed scheduling mechanisms are well-known open problems and are an active area of research; see \cite{Ji:Lin:Shroff:2016} and references therein.   Some distributed algorithms for admission control are the row constraints \cite{Gupta:Musacchio:Walrand:07} \cite{Hamdaoui:Ramanathan:05}, the scaled clique constraints \cite{Gupta:Musacchio:Walrand:07}, and the degree and mixed conditions \cite{Hamdaoui:Ramanathan:05}; the worst-case performance of these sufficient conditions was analyzed in \cite{Ganesan:WN:2014}.     

In the worst case, the row constraints are a factor of $\sigma(G_C)$ away from optimal, where $\sigma(G_C)$ is defined below and is called the \emph{induced star number} of the conflict graph or the \emph{interference degree} \cite{Ganesan:2010} \cite{Chaporkar:Kar:Luo:Sarkar:2008}.   In some cases, for example for conflict graphs constructed from the $2$-hop interference model of the IEEE 802.11 MAC protocol or the $K$-hop interference model, the induced star number $\sigma(G_C)$ can be arbitrarily large \cite{Chaporkar:Kar:Luo:Sarkar:2008}.  For some classes of networks and interference models, the induced star number is bounded from above by a fixed constant \cite{Ganesan:WN:2014} \cite{Chaporkar:Kar:Luo:Sarkar:2008}; for example, the induced star number of a unit disk graph is at most $5$ \cite{Ganesan:WN:2014} and the induced star number of a line graph is at most $2$. 

In the special case where the wireless interference is modeled as primary interference, Shannon's upper bound on the chromatic index of multigraphs \cite{Shannon:1949} gives a distributed admission control algorithm which is a factor of $1.5$ away from optimal. A distributed admission control mechanism was obtained in \cite{Ganesan:WN:2014} and shown to be a factor of $1.25$ away from optimal.  

A \emph{line network} is one where all nodes are positioned in a straight line, say on the $x$-axis.  If each node has a certain radius of coverage and two nodes interfere with each other whenever their coverage areas intersect, then one obtains a conflict graph  called an interval graph \cite{AboElFotoh:etal:2005}.  Conflict graphs arising in spectrum allocation are also interval graphs \cite{Soliman:LG:2016}. Polynomial-time algorithms are known for finding maximum weight independent sets in interval graphs.   

Under the protocol interference model, recent work by Kose et al \cite{Kose:Medard:201711} \cite{Kose:etal:201712} investigates computationally efficient solutions for the scheduling problem.  The problem of finding maximum weight independent sets in claw-free graphs is solvable in polynomial time.  The approach taken in \cite{Kose:etal:201712} to address the situation where the conflict graphs are not claw-free is to add edges to the conflict graph to obtain a claw-free graph.  This preserves the original interference constraints and gives a valid schedule in polynomial time.  Because of the added interference constraints, the throughput is suboptimal.

% 
%====================================
\section{A Distributed Algorithm for the Primary Interference Model} \label{sec:dist:algo:primary:interference}

In this section, a new distributed algorithm for admission control is given and its performance is analyzed.  The focus is on the primary interference model, and it is assumed that each node needs to know the quality-of-service requirements of only neighbors which are at distance at most $d$.  In Section~\ref{sec:dist:algo:preliminaries}, the special cases of $d=0$ and $d=1$ are studied. A distance-$d$ distributed algorithm for admission control is given in Section~\ref{sec:distance:d:distributed} and its performance is analyzed in Section~\ref{sec:distance:d:distr:performance}. Intuition for the worst-case performance bound of the distributed algorithm, its time complexity, and a distributed implementation of the algorithm are given in Section~\ref{sec:intuition:dist:implementation}. 

%====================================
\subsection{Preliminaries} \label{sec:dist:algo:preliminaries}

One model of interference is the primary interference model.  In this model, two links $\ell_i, \ell_j \in L$ in the network graph $G=(V,L)$ cannot be simultaneously active if and only if $\ell_i$ and $\ell_j$ have an endvertex in common.  Equivalently, the conflict graph $G_c$ is the line graph of $G$.  The line graph of a graph $G=(V,L)$, denoted $L(G)$, is the graph with vertex set $L$, and $\ell_i, \ell_j \in L$ are adjacent vertices in the line graph if and only if edges $\ell_i, \ell_j$ are incident to a common vertex in $G$.  

Under the primary interference model, an independent set of vertices in the conflict graph $G_c = L(G)$ correponds to a set of edges in the network graph which forms a matching.  The minimum duration of a schedule satisfying a link demand vector $\tau = (\tau(\ell): \ell \in L)$ is called the \emph{fractional chromatic index} of the edge-weighted graph $(G, \tau)$ and is denoted $T^*(\tau)$ henceforth. 

\begin{Example} \upshape \label{eg:odd:cycle:schedule}
 Suppose the network graph is the odd cycle $C_n = (V,L)$, for some $n \ge 5$, where $L = \{\ell_0, \ell_1, \ldots, \ell_{n-1} \}$.  Suppose the link demand vector is $\tau = (1, 1, \ldots, 1)$.  Under the primary interference model, the maximum number of links which can be simultaneously active is $(n-1)/2$.  Hence, a lower bound for the minimum duration of a schedule satisfying demand $\tau$ is given by $T^*(\tau) \ge \frac{\sum_{\ell \in L} \tau(\ell)}{(n-1)/2} = 2n / (n-1)$.  A schedule whose duration is $2n/(n-1)$ is now constructed. Let $\ell_0, \ell_1, \ldots, \ell_{n-1}$ be the links of the odd cycle in order.  Under primary interference, the set $\{\ell_0, \ell_2, \ell_4, \ldots, \ell_{n-3}\}$ of $(n-1)/2$ links can be simultaneously active.    More generally, $A_i := \{\ell_i, \ell_{i+2}, \ell_{i+4}, \ldots, \ell_{i+n-3} \}$ is a maximal set of links which can be simultaneously active; here, subscripts are taken modulo $n$.  The schedule $t$ defined by $t(A_i) = \frac{2}{n-1}$, $i=0,1,\ldots,n-1$ has total duration $\frac{2n}{n-1}$.  The sum of the cardinalities of the $A_i$'s is $n(n-1)/2$.  Each link is in $(n-1)/2$ of the $A_i$'s.  Hence, each link is active for total duration  $\frac{(n-1)}{2} \frac{2}{(n-1)} = 1$, as desired. \qed 
\end{Example}

Given a network graph $G=(V,L)$ and a link demand vector $\tau = (\tau(\ell): \ell \in L)$, define the degree of $\tau$ at node $v$ by $\delta(\tau, v) := \sum_{\ell: v \sim \ell} \tau(\ell)$, where the sum is over all links which are incident to node $v$. Define the maximum degree $\Delta(\tau) := \max_v \delta(\tau,v)$.  Under primary interference, the time slots assigned to two links which are incident to a common node must be disjoint.  Thus, a lower bound for the duration of an optimal schedule is $T^*(\tau) \ge \Delta(\tau)$. It follows that a necessary condition for demand $\tau$ to be feasible is that the degree $\delta(\tau, v)$ of $\tau$ at every vertex $v$ is at most $1$. 

Shannon studied the problem of color coding wires of electrical units, where devices such as relays and switches are interconnected such that wires coming out of a single point are colored differently. This problem is equivalent to edge-coloring multigraphs. A \emph{multigraph} is a generalization of a graph obtained by allowing more than one (parallel) edge between pairs of vertices.  Shannon gave an upper bound for the chromatic index of multigraphs \cite{Shannon:1949} \cite{Fiorini:Wilson:1977}. 
Lemma~\ref{lem:sufficient:cond:shannon} is a sufficient condition for a link demand vector to be feasible and was obtained in \cite{Ganesan:WN:2014} using Shannon's upper bound.  Below, a new proof from first principles is given which does not use Shannon's upper bound.  
% The nontrivial part of Shannon's proof is for the case where the maximum degree of the multigraph is odd.  Because this part of the proof is not required to deduce Lemma~\ref{lem:sufficient:cond:shannon}, rather than apply Shannon's result, an elementary proof from first principles is given below for Lemma~\ref{lem:sufficient:cond:shannon}. As shown in the proof below, it can be assumed without loss of generality that the multigraphs arising in the admission control problem have even maximum degree. 

\begin{Lemma} \label{lem:sufficient:cond:shannon}
 Let $G=(V,L)$ be a network graph and let $\tau = (\tau(\ell): \ell \in L)$ be a link demand vector.  Under the primary interference model, $\tau$ is feasible within $1$ unit of time if the degree of $\tau$ at each vertex $v$ is at most $2/3$.  This sufficient condition is a factor of at most $1.5$ away from optimal.
\end{Lemma}

\noindent \emph{Proof:}
Because the fraction of time $\tau(\ell)$ a link $\ell$ demands to be active is a rational number, one can divide the time axis into sufficiently small frames that the demand of each link is an integral number of frames.  The time duration of each frame can be further halved so that the demand of a link $\ell$ is an even integer $\mu(\ell)$.  Construct a multigraph $M = (G, \mu)$ by replacing each edge $\ell$ of $G$ by $\mu(\ell)$ parallel edges.  Let $\Delta(M)$ denote the maximum degree of the multigraph $M$. Note that $\Delta(M)$ is even. It can be assumed without loss of generality that $M$ is $\Delta(M)$-regular, for one can add vertices and edges to obtain a $\Delta(M)$-regular multigraph.  Suppose $\Delta(M)=2k$.  By Petersen's $2$-factor theorem, the edge set of $M$ can be decomposed into $k$ $2$-factors.  Each $2$-factor is a union of cycles and hence is $3$-edge-colorable.  Thus, the chromatic index of the multigraph $M$ is at most $3k = 3 \Delta(M)/2$. An edge-coloring of the multigraph yields a valid schedule, and upper bounds on the chromatic index of the multigraph give corresponding upper bounds on the minimum duration of a schedule satisfying demand $\tau$.  Hence, $T^*(\tau) \le \frac{3}{2} \Delta(\tau)$. This proves that a sufficient condition for $\tau$ to be feasible is $\Delta(\tau) \le 2/3$. 

Scaling the resource requirement of $2/3$ in the sufficient condition $\Delta(\tau) \le 2/3$ by a factor of $1.5$ gives the condition $\Delta(\tau) \le 1$, which is a necessary condition for $\tau$ to be feasible.  Hence, the sufficient condition in the assertion is a factor of at most $1.5$ away from optimal. 
\qed

The sufficient condition in Lemma~\ref{lem:sufficient:cond:shannon} performs optimally, for example, when the network graph is a Shannon multigraph, essentially the thick triangle given in the next example.  

\begin{Example} \upshape
 Suppose the network graph $G=(V,L)$ is a $3$-cycle graph and the link demand vector is $\tau = (\frac{1}{3}, \frac{1}{3}, \frac{1}{3})$.  Then, $\Delta(\tau) = \frac{2}{3}$, and so $\tau$ is feasible by Lemma~\ref{lem:sufficient:cond:shannon}. \qed 
\end{Example}

The proof of Lemma~\ref{lem:sufficient:cond:shannon} decomposes the edge set of the network multigraph into $2$-cycles and then uses the fact that each $2$-cycle is $3$-colorable.  The resource requirement of $3$ colors is an overestimate if all the $2$-cycles have even length.  Therefore, in order to investigate the worst-case performance of the above sufficient condition, one can consider bipartite graphs.  In fact, for bipartite graphs $G$, a necessary and sufficient condition for $\tau$ to be feasible is that $\Delta(\tau) \le 1$.   In the simplest case, consider a link demand vector of the form $\tau = (1, 0, \ldots, 0)$: 

\begin{Example} \upshape
Let $G$ be any network graph consisting of links $\ell_1,\ldots,\ell_m$ ($m \ge 1$). Consider the link demand vector $\tau = (1, 0, \ldots, 0)$.  Because $\Delta(\tau) = 1 > \frac{2}{3}$, the distributed admission control algorithm of Lemma~\ref{lem:sufficient:cond:shannon} will conclude that this demand cannot be satisfied.   However, $\tau$ is clearly feasible.

Let $P_S = \{\tau: \Delta(\tau) \le \frac{2}{3} \}$. The smallest $\alpha$ for which $\alpha P_S := \{\tau: \Delta(\tau) \le \alpha \frac{2}{3} \}$ contains $(1, 0, \ldots, 0)$ is $1.5$.  Hence, in the worst case, the sufficient condition in Lemma~\ref{lem:sufficient:cond:shannon} is  a factor of at least $1.5$ away from optimal. Also, $1.5 P_S$ contains $P_I$. Hence, in the worst case, the sufficient condition in Lemma~\ref{lem:sufficient:cond:shannon} is a factor of exactly $1.5$ away from optimal. 
\qed
\end{Example}

A graph $G$ is said to be {\em perfect} if for each induced subgraph $H \subseteq G$, the chromatic number $\chi(H)$ and clique number $\omega(H)$ are equal.  An odd hole in a graph is an induced odd cycle of length at least $5$. An odd antihole in a graph is an induced subgraph which is the complement of an odd cycle of length at least $5$.  The strong perfect graph theorem asserts that a graph is perfect if and only if it does not contain any odd holes or odd antiholes.  Given a conflict graph $G_c=(L,L')$ and demand $\tau = (\tau(\ell): \ell \in L)$, a necessary condition for $\tau$ to be feasible is that $\tau(K) \le 1$ for each clique $K$ in the conflict graph, where $\tau(K) := \sum_{\ell \in K} \tau(\ell)$.   This necessary condition, called the \emph{clique constraints}, can be scaled to give sufficient conditions for admission control  \cite{Gupta:Musacchio:Walrand:07}, \cite{Ganesan:WN:2014}.   Let $P_{\mbox{clique}} = P_{\mbox{clique}}(G_C)$ denote $\{\tau: \tau(K) \le 1, \mbox{ for all cliques } K \mbox{ in } G_C\}$.  Then, $P_I = P_{\mbox{clique}}$ if and only if $G_C$ is a perfect graph \cite{Grotschel:Lovasz:Schrijver:1988}.   Under primary interference, the following result gives a sufficient condition for admission control.  

\begin{Lemma} \label{lem:suff:cond:0.8}  \cite{Ganesan:2009} \cite{Ganesan:WN:2014} \cite{Ganesan:2008} 
Let $G=(V,L)$ be a network graph and let $\tau = (\tau(\ell): \ell \in L)$ be a link demand vector. Then, under the primary interference model, $\tau$ is feasible if the degree of $\tau$ at each vertex is at most $0.8$ and the sum of the demands of links in every triangle in $G$ is at most $0.8$.  In the worst case, this sufficient condition is a factor of  $1.25$ away from optimal.
\end{Lemma}

\noindent \emph{Proof:}
Under the primary interference model, the conflict graph $G_c$ is the line graph $L(G)$.  The imperfection ratio of $G_c$, denoted $\imp(G_c)$, is defined to be $\imp(G_c) := \sup_{ \tau \ne 0} \frac{\chi_f(G_c, \tau)}{\omega(G_c, \tau)}$, where $\omega(G_c, \tau)$ denotes the clique number of the vertex-weighted graph $(G_c, \tau)$, and the supremum is taken over all nonzero integral vectors $\tau$. If $G_c$ has no odd holes, then $\imp(G_c) = 1$, and if the minimum length of an odd hole in $G_c$ is $g \ge 5$, then $\imp(G_c) = \frac{g}{g-1}$ \cite{Gerke:McDiarmid:2001}.  Because $g \ge 5$, one has $\imp(G_c) \le \frac{5}{4}$ and $\omega(G_c, \tau) \le \chi_f(G_c, \tau) \le 1.25 \omega(G_c, \tau)$.  The following sufficient condition for admission control is thus obtained: $\tau$ is feasible if $\omega(G_c, \tau) \le 0.8$.  Each clique in the line graph $G_c = L(G)$ corresponds either to a set of links in $G$ which are incident to a common node in $G$ or to a set of links which forms a triangle in $G$. This gives the sufficient condition in the assertion. It is clear this sufficient condition is a factor of at most $1.25$ away from optimal. 

Consider the $4$-cycle network graph $G$ with link demand vector $\tau = (\frac{1}{2}, \frac{1}{2}, \frac{1}{2}, \frac{1}{2})$.  Because $\Delta(\tau) = 1 > 0.8$, the admission control protocol denies admission to $\tau$ even though $\tau$ is feasible. Let $P_S$ denote the set of demands for the said network graph which satisfy the sufficient condition in the assertion.  The smallest value of $\alpha$ for which $\alpha P_S$ contains $\tau$ is $1.25$. Hence, in the worst-case, the sufficient condition is a factor of at least $1.25$ away from optimal.  Scaling the sufficient condition by a factor of $1.25$ gives a necessary condition.  
\qed

Other network graphs $G$ for which the sufficient condition above exhibits its worst-case performance are those graphs $G$ for which the line graph $L(G)$ is perfect, because in such cases $\omega(G_c, \tau) = \chi_f(G_c, \tau)$. Trotter \cite{Trotter:1977} showed that line graphs are perfect if and only if they do not contain odd holes. This implies that if the conflict graph does not contain any induced odd cycles of length at least $5$, or equivalently, if the network graph does not contain odd cycles of length at least $5$, then one obtains a sufficient condition for admission control which is also necessary by replacing the $0.8$'s in Lemma~\ref{lem:suff:cond:0.8} with $1$'s.  The factors $0.8$ in the sufficient condition are due to the possibility of induced $5$-cycles in $G_c$.  Hence, the sufficient condition above performs optimally if one takes $G$ to be a $5$-cycle graph with demand $\tau = (0.4, \ldots, 0.4)$. 

The worst-case performance of the above sufficient condition is $\frac{5}{4}$.  This bottleneck on performance is due to the possible existence of $5$-cycles in the network graph which are not taken into consideration by the sufficient condition in Lemma~\ref{lem:sufficient:cond:shannon} - only up to $3$-cycles in the network graph are considered when estimating the resource requirements, and so a conservative factor of $0.8$ is included to account for any $5$-cycles which might exist in the network graph. 
One way to improve the performance of the distributed algorithm is to increase the amount of global information available at each node in the network graph.  This is the approach taken in Section~\ref{sec:distance:d:distributed}. 

%===============================
\subsection{A Distance-$d$ Distributed Algorithm} \label{sec:distance:d:distributed}

A distance-$d$ distributed algorithm is one which assumes that for each node $v \in V$ in the network graph $G=(V,L)$, node $v$ knows the demands of all communication links incident to it and the demands of all communication links $\ell=(x,y)$ between nodes $x$ and $y$ whenever the link's endpoints $x$ and $y$ are at distance at most $d$ from $v$, and each node $v$ has no further global information.   In this case, it is said that the {\em degree of centralization} is $d$. Thus, if the degree of centralization $d$ of a distributed algorithm is $1$, then each node knows the demands of all links incident to it and the demands of all links between its neighbors (cf. Lemma~\ref{lem:suff:cond:0.8}).   In the special case when $d=0$, it is assumed that each node knows only the demands of all links incident to it; this special case was studied in Section~\ref{sec:dist:algo:preliminaries} (see Lemma~\ref{lem:sufficient:cond:shannon}), and so in the rest of this section it is assumed that $d \ge 1$.  

A formal definition of the distance-$d$ distributed algorithm for admission control is as follows. Let $G=(V,L)$ be a network graph and let $v \in V$.  Let $G_i(v)$ denote the set of vertices in $G$ whose distance to $v$ is exactly $i$.  Thus, $G_0(v) = \{v \}$ and $G_1(v)$ is the set of neighbors of $v$.  The subsets $G_0(v), G_1(v), \ldots$ are called the layers of the distance partition of $G$ with respect to $v$.  Define the ball of radius $d$ centered at $v$ by
$$ W_{v,d} = \{v\} \cup G_1(v) \cup \cdots \cup G_d(v) ~~(d \ge 1).$$
Let $G_{v,d}$ denote the subgraph of $G$ induced by $W_{v,d}$. Let $\tau = (\tau(\ell): \ell \in L)$ be a link demand vector.  Let $T^*(G_{v,d}, \tau)$ denote the minimum duration of a schedule satisfying the demands of all links in the induced subgraph $G_{v,d}$.  Because such a schedule considers the demands of only links which are in the ball of radius $d$ centered at node $v$, its duration is a lower bound on $T^*(\tau)$.  Note that $T^*(G_{v,d}, \tau)$ does not depend on all of $\tau$ but only on the demands of those links which are in the distance-$d$ neighborhood $G_{v,d}$ of node $v$.  Define $T_d^*(\tau) := \max_{v \in V} T^*(G_{v,d}, \tau)$.  Then, $T_d^*(\tau)$ is a lower bound for $T^*(\tau)$ and is a nondecreasing function of $d$.  

The following is a distance-$d$ distributed algorithm for admission control: each node $v$ computes $T^*(G_{v,d}, \tau)$ based on demands of all links which are in the ball of radius $d$ centered at $v$.  If this quantity is less than some threshold (to be defined below), then the flows will be accepted; otherwise, the flows will be denied admission. The performance of this distributed algorithm is analyzed next.

\subsection{Performance Analysis} \label{sec:distance:d:distr:performance}

\begin{Theorem} \label{thm:dist:d:suff:cond:perf}
 Let $G=(V,L)$ be a network graph, let $\tau = (\tau(\ell): \ell \in \L)$ be a link demand vector, and let $d \ge 1$.  Let $T^*(G_{v,d}, \tau)$ denote the minimum duration of a schedule which satisfies the demands of all links in the distance-$d$ neighborhood graph $G_{v,d}$ of node $v$.  Let $\alpha_d = \frac{2d+3}{2d+2}$.  A sufficient condition for $\tau$ to be feasible is that $T^*(G_{v,d}, \tau) \le \frac{1}{\alpha_d}$ for each node $v$.  This sufficient condition is a factor of at most $\alpha_d$ away from optimal.
\end{Theorem}

\noindent \emph{Proof:}
For the first part of the proof, it suffices to show that $\frac{T^*(\tau)}{T_d^*(\tau)} \le \alpha_d$, where the numerator $T^*(\tau)$ is the fractional chromatic index of the edge-weighted graph $(G, \tau)$ and $T^*_d(\tau) = \max_{v \in V} T^*(G_{v,d}, \tau)$; this would imply that any $\tau$ satisfying the sufficient condition has $T^*(\tau) \le T_d^*(\tau) \alpha_d \le \frac{1}{\alpha_d} \alpha_d = 1$.  Let $W$ be a subset of $V$ of odd cardinality $k$, and let $E(G[W])$ denote the set of all links of $G$ in the induced subgraph $G[W]$.  Define $$\Lambda(\tau,W) := \frac{\sum_{\ell \in E(G[W])}\tau(\ell)}{(k-1)/2}.$$  Because the maximum size of a matching in the induced subgraph $G[W]$ is at most $(k-1)/2$, $\Lambda(\tau,W) \le T^*(\tau)$.  Define $\Lambda(\tau) := \max_{W \subseteq V} \Lambda(\tau,W)$, where the maximum is over all subsets $W \subseteq V$ of odd cardinality. It is clear that $\Lambda(\tau) \le T^*(\tau)$.  It follows from Edmonds' Theorem (cf. \cite{Scheinerman:Ullman:1997} \cite{Edmonds:1965}) that $T^*(\tau) = \max\{\Delta(\tau), \Lambda(\tau) \}$. 

It will now be proved that $\Delta(\tau) \le \alpha_d T_d^*(\tau)$ and $\Lambda(\tau) \le \alpha_d T_d^*(\tau)$; this would imply $T^*(\tau) \le \alpha_d T_d^*(\tau)$.   The minimum duration of a schedule satisfying the demands of all links in the distance-$d$ neighborhood $G_{v,d}$ of node $v$ is at least the degree $\delta(\tau, v)$ of $\tau$ at node $v$. Hence, $\Delta(\tau) \le T_d^*(\tau)$, which is at most $\alpha_d T_d^*(\tau)$ because $\alpha_d \ge 1$.  The inequality $\Lambda(\tau) \le \alpha_d T_d^*(\tau)$ will be proved next.  Fix $k$ such that $3 \le k \le |V(G)|$ and $k$ is odd.  Let $W$ be a subset of $V(G)$ of cardinality $k$.  It suffices to show that 
$$\frac{\sum_{\ell \in E(G[W])} \tau(\ell)}{(k-1)/2} \le \alpha_d T_d^*(\tau).$$

Consider two cases for $k$ (this proof technique is from \cite{Gerke:McDiarmid:2001}).  First suppose that $3 \le k < 2d+3$.  The maximum size of a matching in $G[W]$ is at most $(k-1)/2$, whence the minimum duration of a schedule satisfying the link demand vector $(\tau(\ell): \ell \in E(G[W]))$ is at least $\frac{1}{(k-1)/2} \sum_{\ell \in E(G[W])} \tau(\ell)$.   Because $k \le 2d+1$, each connected component in the induced subgraph $G[W]$ is contained in some ball of radius $d$ in $G$.  Hence, the minimum duration of a schedule satisfying the link demand vector $(\tau(\ell): \ell \in E(G[W]))$ is at most $T_d^*(\tau)$.  It follows that
$$\frac{\sum_{\ell \in E(G[W])} \tau(\ell)}{(k-1)/2} \le T_d^*(\tau) \le \alpha_d T_d^*(\tau).$$

Now suppose $k \ge 2d+3$, where $|W| = k$ again.  One can double-count the sum of demands of all links in $E(G[W])$ (this is essentially the handshaking theorem) to obtain that $2 \sum_{\ell \in E(G[W])} \tau(\ell) \le k \Delta(\tau)$.   Hence,
$$ \frac{\sum_{\ell \in E(G[W])} \tau(\ell)}{(k-1)/2} \le \frac{(k/2) \Delta(\tau)}{(k-1)/2}  \le \frac{k}{k-1} T_d^*(\tau) \le \alpha_d T_d^*(\tau), $$
where the last two inequalities follow from the fact that $\Delta(\tau) \le T_d^*(\tau)$ and that $k/(k-1)$ is a decreasing function of $k$. This proves that $\Lambda(\tau) \le \alpha_d T_d^*(\tau)$, as required.

Thus, for any given network graph $G=(V,E)$, link demand vector $\tau$, and distance $d \ge 1$, one has $T^*(\tau) \le \alpha_d T_d^*(\tau)$.  In particular, if $T_d^*(\tau) \le \frac{1}{\alpha_d}$, then $T^*(\tau) \le 1$.  This gives the sufficient condition for admission control in the assertion in which each node uses localized information up to distance $d$ in the network graph.   Scaling the right-hand side of the sufficient condition $T^*(G_{v,d}, \tau) \le \frac{1}{\alpha_d}$ by a factor of $\alpha_d$ gives $T^*(G_{v,d}, \tau) \le 1$, which is a necessary condition for $\tau$ to be feasible. Hence, the sufficient condition in the assertion is a factor of at most $\alpha_d$ away from optimal.
\qed

The bound $\alpha_d$ on the performance of the above sufficient condition is best possible:

\begin{Proposition}
 In the worst case, the sufficient condition in Theorem~\ref{thm:dist:d:suff:cond:perf} is a factor of exactly $\alpha_d$ away from optimal.
\end{Proposition}

\noindent \emph{Proof:}
Let $G$ be any network graph and let $d \ge 1$. Let $P_S = \{\tau: T_d^*(\tau) \le \frac{1}{\alpha_d} \}$  be the set of all link demand vectors which satisfy the sufficient condition in Theorem~\ref{thm:dist:d:suff:cond:perf}.  The link demand vector $\tau = (1, 0, \ldots, 0)$ is feasible but is not contained in $P_S$. The smallest value of $\alpha$ for which $\alpha P_S$ contains $\tau$ is $\alpha_d$.  This proves that in the worst case, the sufficient condition in Theorem~\ref{thm:dist:d:suff:cond:perf} is a factor of at least $\alpha_d$ away from optimal. By Theorem~\ref{thm:dist:d:suff:cond:perf}, the sufficient condition is also a factor of at most $\alpha_d$ away from optimal.
\qed

In situations where battery life is important and communication with distant nodes should be avoided, distributed mechanisms are preferable.  The fundamental question ``what are the limits to the performance that is achievable if each node has information only up to $d$ hops away'' has been answered for the case of the primary interference model: in the worst case, the distance-$d$ distributed algorithm would overestimate the amount of resources required to satisfy the given QoS bandwidth requirements by a factor of $\frac{2d+3}{2d+2}$.  As $d$ increases, there is an increase in communication cost because information from more distant nodes needs to be communicated to each node.  As $d$ increases, the local processing cost at each node for solving  the admission control problem with the available information also increases.   This latter problem can be solved in polynomial time in the case of primary interference model.  Between the two types of cost, it is the former which is usually considered more expensive in the field of distributed algorithms; nonetheless, a brief discussion of the local processing cost is given in the next.  

%===============================================================
\subsection{Complexity Analysis and Distributed Algorithm} \label{sec:intuition:dist:implementation}

In Section~\ref{sec:distance:d:distr:performance}, a sufficient condition for admission control, namely a distance-$d$ distributed algorithm, was proposed and its worst-case performance was analyzed. In the present subsection, intuition for the threshold $\alpha_d = (2d+3)/(2d+2)$ in the distributed algorithm is given, the time complexity of the distributed algorithm and the performance-complexity tradeoff are analyzed, and a distributed implementation of the sufficient condition is given in algorithmic format. 

\textbf{Intuition:} The intuition for the threshold $\frac{1}{\alpha_d}$ in Theorem~\ref{thm:dist:d:suff:cond:perf} can be understood by considering large odd cycles with uniform demand patterns, as follows.  Theorem~\ref{thm:dist:d:suff:cond:perf} states that under the primary interference model, a sufficient condition for $\tau$ to be feasible is that the resource estimate $T^*(G_{v,d}, \tau)$ computed by each node $v$, based on demands in its distance-$d$ neighborhood subgraph $G_{v,d}$, is at most $\frac{1}{\alpha_d}$, where $\alpha_d = \frac{2d+3}{2d+2}$. Recall that under the primary interference model, the conflict graph $G_c$ is the line graph $L(G)$ of the network graph $G=(V,L)$. If the conflict graph $G_c$ is a perfect graph, then the clique constraints are optimal. Cliques in a line graph $G_c=L(G)$ correspond to triangles $K_3$ or stars $K_{1,r}$ in the network graph $G$, and so are accounted for in the resource estimate $T^*(\tau,G_{v,d})$.  It is known that a line graph $L(G)$ is imperfect if and only if it contains an odd hole, and such an odd hole must of course be induced by an odd cycle in $G$. Thus, a bottleneck in the performance of distance-$d$ distributed algorithms is the existence of odd cycles in the network graph $G$ which are not contained in any distance-$d$ neighborhood subgraph $G_{v,d}$. 

Each odd cycle in $G$ of length at most $2d+1$ is contained in some distance-$d$ neighborhood subgraph $G_{v,d}$, whereas the odd cycles of length at least $2d+3$ are not contained in any of these subgraphs.  The imperfection ratio of an odd cycle $C_n$ is $n/(n-1)$, and so the smaller odd cycles have a larger imperfection ratio.  In the worst case, an adversary who knows the exact value of $d$ used by the distance-$d$ distributed algorithm can choose a network graph $G$ which contains a cycle of length exactly $2d+3$.  

Suppose that the network graph $G$ is the single large cycle $C_{2d+3}$. Suppose the distance-$d$ distributed algorithm is used as a sufficient condition for admission control, with uniform demand $\tau = (\frac{1}{2}, \ldots, \frac{1}{2})$. The distance-$d$ neighborhood subgraph $G_{v,d}$ centered at each node $v$ is bipartite and hence does not contain the entire cycle. Consequently, the local estimate (i.e. the estimate of resource requirement, computed using localized information) is $T_d^*(\tau) = 1$.  Example~\ref{eg:odd:cycle:schedule} proves that the exact value of the resource requirement is $T^*(\tau) = \frac{2d+3}{2d+2}$, which is $\alpha_d$.  This example of a large, odd cycle with uniform demand pattern shows that in order to obtain an estimate of the actual resource requirement $T^*(\tau)$, one must multiply the local estimate $T_d^*(\tau)$ by a factor of at least $\alpha_d$. However, doing so can sometimes overestimate the resource requirement by this factor $\alpha_d$, as can be seen by considering the example where the demand vector is $\tau = (\frac{1}{2}, \ldots, \frac{1}{2}, 0)$, for which $T_d^*(\tau) = T^*(\tau) = 1$. 

The threshold $\alpha_d$ depends only on topology information, i.e. on the exact value of the parameter $d$, and is the same for all nodes in the network.  This threshold does not depend on or use information about the localized demand information that is also available to each node.  An interesting problem is to determine how each node can use its local demand pattern to rule out the possibility of the worst-case situation of uniform demand pattern mentioned above and compute a better threshold. 

\textbf{Complexity analysis and performance-complexity tradeoff:} The complexity of the distance-$d$ distributed algorithm and the performance-complexity tradeoff are analyzed next. It was seen that as $d$ increases, the performance of the distributed algorithm improves to $(2d+3)/(2d+2)$ because more global information is available at each node. However, as $d$ increases, the complexity of the local algorithm at each node $v$ also increases - it will be seen that the time complexity of the distance-$d$ distributed algorithm executed at each node $v$ to compute its local estimate is $O(\Delta^{5d})$, where $\Delta$ is the maximum degree of a vertex in the graph.

Each node $v$ in the network graph $G$ needs to compute the local estimate $T^*(G_{v,d}, \tau)$, which is defined to be the minimum duration of a schedule that satisfies the demands of all links its its distance-$d$ neighborhood subgraph $G_{v,d}$.  Recall from the proof of Theorem~\ref{thm:dist:d:suff:cond:perf} that it follows from Edmonds' matching polytope theorem that the fractional chromatic index of a weighted graph $(G, \tau)$ is equal to   $T^*(\tau) = \max\{\Delta(\tau), \Lambda(\tau) \}$; however, a node $v$ cannot use this formula directly on its subgraph to compute its local estimate in polynomial time because the density $\Lambda(\tau)$ depends on an exponential number of odd subsets.

Each node $v$ needs to compute the fractional chromatic index of the edge-weighted, distance-$d$ neighborhood subgraph $(G_{v,d}, \tau)$, and this value can be defined by the linear program $(\min 1^T t$ subject to $Mt \ge \tau, t \ge 0$), where $M$ is the edge-matching incidence matrix of the subgraph. The number of variables in the linear program is equal to the number of matchings in the subgraph, which can be exponential in the size of the input to node $v$.  The dual linear program is ($\max y^T \tau$ subject to $y^T M \le 1, y \ge 0$), and while it can have an exponential number of constraints, the separation problem of finding a violating constraint can be solved in polynomial time by finding a maximum weighted matching in the subgraph \cite{Grotschel:Lovasz:Schrijver:1988}\cite{Schrijver:2003}.  Recall that the ellipsoid method can be used to even solve large linear programs in polynomial time if the separation problem can be solved in polynomial time (as is the case in the linear program of interest here).  Hence, each node $v$ can compute its local estimate $T^*(G_{v,d}, \tau)$ in time polynomial in the size of its input.

A strongly polynomial time algorithm for computing the fractional chromatic index of an edge-weighted graph is given in \cite{Hajek:Sasaki:1988}. Using this algorithm, each node $v$ can compute the local estimate $T^*(G_{v,d}, \tau)$ in time $O(n^5)$, where $n$ is the number of nodes in the distance-$d$ neighborhood subgraph $G_{v,d}$.  The number of nodes in a distance-$d$ neighborhood can be bounded from above by $1+\Delta+\Delta (\Delta-1) + \cdots + \Delta (\Delta-1)^{d-1}= O(\Delta^d)$, where $\Delta$ denotes the maximum degree of a vertex in the subgraph.  Hence, each node $v$ can compute its local estimate $T^*(G_{v,d}, \tau)$ in $O(\Delta^{5d})$ time.  The performance-complexity tradeoff, as a function of the parameter $d$, is summarized by the following result.

\begin{Theorem}
The complexity of the sufficient condition of Theorem~\ref{thm:dist:d:suff:cond:perf}, i.e. the time required by each node $v$ of the distance-$d$ distributed algorithm to compute its local estimate $T^*(G_{v,d},\tau)$, is $O(\Delta^{5d})$, where $\Delta$ is the maximum degree of a vertex in the network graph $G=(V,L)$, and the performance of the distance-$d$ distributed algorithm is away from optimal by a factor of at most $(2d+3)/(2d+2)$.  
\end{Theorem}

Recently, a strongly polynomial time algorithm was given for computing not just the fractional chromatic index $\max\{\Delta(\tau), \Lambda(\tau)\}$, but also for the density $\Lambda(\tau)$, for any edge-weighted multigraph with rational weights \cite{Chen:Zang:Zhao:2019:SIAM}. Their algorithm runs in time $O(mn + n^5 \ell^2 \log(n^2/\ell))$, where $n$, $m$, and $\ell$ denote the number of vertices, number of edges in the multigraph, and number of edges in the underlying graph, respectively. %When the multigraph is a simple graph, $m = \ell$. 

\textbf{Distributed algorithm:} A distributed algorithm which implements the sufficient condition of Theorem~\ref{thm:dist:d:suff:cond:perf} is given below in an algorithmic format. It is assumed that each node $v$ in the network graph $G=(V,L)$ knows the distance-$d$ neighborhood subgraph $G_{v,d}$ and the demands of all links in this neighborhood. Recall that a sufficient condition for a link demand vector $\tau$ to be feasible is that the local estimate $T^*(G_{v,d}, \tau)$ computed at each node $v$ is at most $1/\alpha_d = (2d+2)/(2d+3)$. 

If the local estimate computed at a node $v$ is larger than $1/\alpha_d$, then this negative result could be due to the high demand of any link in the neighborhood $G_{v,d}$. A conservative implementation of the sufficient condition is to deny admission to all flows in the neighborhood $G_{v,d}$ if the local estimate at node $v$ exceeds $1/\alpha_d$. Thus, a flow $\tau(xy)$ between nodes $x$ and $y$ is admitted if and only if, for each node $v$ for which $G_{v,d}$ contains the link $xy$, the local estimate computed at $v$ is at most $1/\alpha_d$.  

A distributed algorithm based on the above idea is shown in Algorithm~\ref{algo:dist:d}. 
After a node $v$ has computed its local estimate $T^*(G_{v,d}, \tau)$, referred to by the variable \verb|T| in Algorithm~\ref{algo:dist:d}, node $v$ sends a message to all nodes in its distance-$d$ neighborhood: node $v$ sends the message \verb|(v,feasible)| if the local estimate is at most $1/\alpha_d$, and it sends the message \verb|(v,infeasible)| if the local estimate is larger than $1/\alpha_d$.   This transmission to all nodes in the distance-$d$ neighborhood subgraph can be achieved by a distance-$d$ flooding algorithm \cite{Peleg:2000}.  This is the usual flooding algorithm, but with an extra counter variable in each message which has initial value $d$; a node which receives a message on a link decrements this counter by $1$ and forwards the message on all other links if this counter is positive.  If nodes $x$ and $y$ both receive the message \verb|(v,infeasible)| from any node $v$ in their distance-$d$ neighborhood, they conclude (possibly too conservatively) that the demand $\tau(xy)$ is infeasible and so this flow is denied admission.

\begin{Algorithm} \upshape \label{algo:dist:d} \textbf{dist\_dDistrAdmissionControl} 
\begin{verbatim}
For each vertex v of network graph:
    compute the local estimate T
    if T <= (2d+2)/(2d+3):
        send (v, feasible) to all
        nodes in d-hop neighborhood
    else:
        send (v, infeasible) to all
        nodes in d-hop neighborhood

For each link xy of network graph:    
    if both x and y received a
    message (v, infeasible) from
    some node v: 
        reject flow xy
    else:
        accept flow xy
\end{verbatim}
\end{Algorithm}

Notice that each node or link in Algorithm~\ref{algo:dist:d} relies only on localized information to make its decision, and hence the algorithm can be implemented in a distributed fashion. The algorithm takes a conservative approach because a link's demand can influence nodes up to $2d$ hops away: it is possible for a flow $xy$ to be denied admission because of an infeasibility result from a node $v$ that is $d$ hops away from $xy$, and node $v$'s infeasibility result could in turn be due to the high demand of a link $ab$ that is $d$ hops away from $v$. Even though links $ab$ and $xy$ are in the same distance-$d$ neighborhood of node $v$, they might be about $2d$ hops away from each other.
%===============================================================

\section{Performance of Row Constraints in Line Networks Under the Protocol Interference Model} \label{sec:row:line:networks:protocol:interference:model}

A simple, distributed mechanism for both admission control and scheduling is given by the {\em row constraints}, which can be obtained by generalizing the greedy graph coloring algorithm to weighted graphs, as follows.   Let $G_c = (L, L')$ be a conflict graph and let $\tau = (\tau(\ell): \ell \in L)$ be a link demand vector.  A sufficient condition for $\tau$ to be feasible within duration $T$ is that $\tau(\ell) + \tau(\Gamma(\ell)) \le T$, for all $\ell \in L$, where $\Gamma(\ell)$ denotes the set of neighbors in $G_c$ of vertex $\ell$, and $\tau(\Gamma(\ell)): = \sum_{\ell' \in \Gamma(\ell)} \tau(\ell')$. This sufficient condition is called the {\em row constraints} (cf. \cite{Gupta:Musacchio:Walrand:07} \cite{Hamdaoui:Ramanathan:05} \cite{Wu:Srikant:2006}).  If $\tau$ is feasible, then a feasible schedule can be obtained by allocating any time interval to $\ell$ which is disjoint from the time intervals allocated to its neighbors in $G_c$.

While the problem of computing the exact value of the fractional chromatic number $T^*(G_c, \tau)$ is NP-hard \cite{Grotschel:Lovasz:Schrijver:1981}, the quantity $\max_{\ell \in L} \{ \tau(\ell) + \tau(\Gamma(\ell)) \}$ can be computed efficiently and is an upper bound on $T^*(G_c, \tau)$. 
The worst-case performance of the row constraints is the largest possible factor by which the row constraints can overestimate the resources required for accepting demand $\tau$, and is equal to $\sup_{\tau \in P_I} \max_{\ell \in V(G_C)} \{ \tau(\ell) + \tau(\Gamma(\ell)) \}$.   A simple formula for this expression was obtained in \cite{Chaporkar:Kar:Luo:Sarkar:2008} and independently in \cite{Ganesan:2008} \cite{Ganesan:2009} \cite{Ganesan:2010}. This result is recalled in Lemma~\ref{lem:AML} below, which says that in the worst case, the row constraints are a factor of $\sigma(G_c)$ away from optimal, where the graph invariant $\sigma(G_c)$, defined below, is called the induced star number of $G_c$. Thus, it is desired that this quantity be as small as possible.

In Section~\ref{sec:line:networks:preliminaries}, the definition of the induced star number of a graph and a result on the worst-case performance of the row constraints are recalled.  The protocol interference model is also described.  In Section~\ref{sec:line:networks:perf:analysis}, it is shown that for line networks under the protocol interference model, the row constraints are a factor of at most $3$ away from optimal.  Further, it is shown that this bound is best possible. 

%-------------------------------------------------------
\subsection{Preliminaries} \label{sec:line:networks:preliminaries}

In this section, results on the induced star number of a graph and the protocol interference model are recalled.  The \emph{induced star number} of a graph $G$, denoted by $\sigma(G)$, is the 
number of leaf vertices in the maximum sized induced star subgraph of $G$, i.e. 
$$\sigma(G) := \max_{v \in V} \alpha(G[G_1(v)]),$$ where $\alpha(G)$ denotes 
the independence number of $G$, $G[W]$ denotes the subgraph of $G$ induced 
by 
$W \subseteq V$, and $G_1(v)$ denotes the set of neighbors in $G$ of vertex $v$. The induced star number of the conflict graph, dentoed by $\sigma(G_c)$, is referred to in \cite{Chaporkar:Kar:Luo:Sarkar:2008} as the  {\em interference degree} of the network and is the maximum number of pairwise noninterfering links in $L$ which interfere with a common link.  

\begin{Lemma} \label{lem:AML} \cite{Chaporkar:Kar:Luo:Sarkar:2008}  \cite{Ganesan:2009} \cite{Ganesan:2008}  \cite{Ganesan:2010} 
 Let $G_c = (L, L')$ be any graph.  Then,
 $$ \sup_{\tau \ne 0} \frac{\max_{v \in V(G_c)} \{ \tau(\ell) + \tau(\Gamma(\ell)) \}}{T^*(G_c, \tau)}  = \sigma(G_c).$$
\end{Lemma}

Lemma~\ref{lem:AML} says that in the worst case, the row constraints are a factor of at most $\sigma(G_C)$ away from optimal.  

For some classes of networks and interference models such as tree networks under the protocol interference model \cite[Theorem~3]{Kose:Medard:201711} and Bluetooth and sensor networks \cite[Example~5]{Hamdaoui:Ramanathan:05}, the conflict graph $G_C$ is a disjoint union of complete graphs, i.e. $\sigma(G_C) = 1$.   The graph $K_{1,3}$ is called a claw.  A graph is claw-free if it does not contain a claw as an induced subgraph. Equivalently, $G_C$ is claw-free if and only if $\sigma(G_C) \le 2$.  The induced star number of a line graph is at most $2$, and so every line graph is claw-free.  Claw-free graphs were initially studied as a generalization of line graphs.

%================================================

%todo: look at gupta kumar paper on this model

The protocol interference model is described next.  Let $V$ be the set of nodes of a wireless network.  Assume that each node has 
the same transmission range $r_T$.  Let $N(i)$ denote the set of nodes within 
transmission range of node $i \in V$.  A (multicast) transmission is defined to 
be a pair $(i,J)$ where $i \in V$ is called the transmitter and $J \subseteq V$ 
is a set of receivers.  A transmission $(i,J)$ is said to be \emph{valid} if $J 
\subseteq N(i)$, i.e. if each of the receivers is within the radius of coverage 
of the transmitter. 

%for a graph, let E(G) denote the edge set of G

%give an example of transmissions. say that for simplicity of exposition, if 
% J_k has only one node A, then we write (i_k,A) for the transmission (i_k,{A})

Given a set $\{(i_k, J_k): k \in K\}$ of valid transmissions over some index set 
$K$, suppose some pairs of transmissions cannot be scheduled simultaneously due to 
wireless interference.   One can construct a conflict graph $G_c$ to capture 
which transmissions interference with each other.  Each transmission corresponds 
to a vertex in the conflict graph $G_c$, and two vertices are adjacent in the 
conflict graph $G_c$ if and only if the corresponding two transmissions interfere 
with each other. Assume the positions of the nodes are known, so that the 
distance $\mbox{dist}(i,j)$ between any pair of nodes $i,j \in V$ can be 
computed.  

\begin{Construction} \upshape \label{construction:KoseMedard} \textbf{Conflict graph of a network under the protocol interference model.} 
Given a set $V$ of wireless nodes in the Euclidean plane and a set $\{(i_k,J_k): k \in K\}$ of valid transmissions, the conflict graph $G_c$ is constructed as follows.  The vertex set of $G_c$ is the set 
of valid transmissions; denote the transmission $(i_k, J_k)$ by the vertex $v_k 
\in V(G_c)$.  Two vertices $v_1 = (k_1, J_1)$ and $v_2 = (i_2, J_2)$ are defined 
to be adjacent in $G_c$ if and only if any one of the following conditions is 
satisfied:

\bigskip \noindent 
(a) $i_1 = i_2$ \\
(b) $i_1 \in J_2$ or $i_2 \in J_1$ \\
(c) $J_1 \cap J_2 \ne \phi$ \\
(d) $\mbox{dist}(i_2,j) < \mbox{dist}(i_1,j)$ for some $j \in J_1$ \\
(e) $\mbox{dist}(i_1,j) < \mbox{dist}(i_2,j)$ for some $j \in J_2$.   
\qed
\end{Construction}

\bigskip \noindent  Implicit in this construction of the conflict graph is the 
following model of interference, usually referred to as the protocol interference model (cf. \cite{Gupta:Kumar:2000}).  Condition (a) says that a node has at most one 
transmitter, and (b) says that a node cannot transmit and receive at the same 
time.  Condition (c) models the constraint that a node has at most one 
receiver.   In order for a transmission $(i_1,J_1)$ to be successful, each 
receiver in $J_1$ must be closer to transmitter $i_1$ than to any other 
transmitter.  If any receiving node in $J_1$  is closer to another 
transmitter $i_2$, then the interference is considered to be intolerable and the 
transmissions $(i_1,J_1)$ and $(i_2,J_2)$ are said to interfere with each other.  
This conflict is captured by condition (d). Similarly, condition (e) captures 
the interference experienced by the receivers in $J_2$.

In the next section, it is shown that $\sigma(G_c) \le 3$ for certain classes of networks.  These are straight line networks for which the demands may be for unicast or multicast transmissions, and the protocol interference model defines which links interfere with each other.  The conflict graph of line networks under the protocol interference model has been studied recently in K\"ose and M\'edard 
\cite{Kose:Medard:201711} and K\"ose et al. \cite{Kose:etal:201712}.  
The proof of the result in \cite{Kose:Medard:201711} \cite{Kose:etal:201712} that the conflict graphs arising in this context are claw-free seems to be incorrect; as proved in 
Theorem~\ref{thm:countereg:KoseMedard} below, there exists a line network for which the conflict graph contains a claw.

%===========================================

\subsection{Performance Analysis} \label{sec:line:networks:perf:analysis}

In this section, the focus is on {\em line networks}, which are networks satisfying the 
condition that all the wireless nodes lie on the same line, say on the $x$-axis.  Recall that the row constraints give a sufficient condition for distributed admission control.  In this section, it is shown that given a line network, if the interference is modeled by the protocol interference model, then the row constraints are a factor of at most $3$ away from optimal (cf. Theorem~\ref{thm:line:networks:sigma:le:3}).

Denote the position of node $A \in V$ by $x_{pos}(A)$.   The closed 
interval $\{x: a \le x \le b\}$ on the real number line is denoted by $[a,b]$.  
In this section, it is assumed that sinks are placed to the right of the source, so 
that information travels towards the right (eastward) and hence each 
transmission $(i_k,J_k)$ satisfies the property that the $x$-coordinate of node 
$i_k$ is at most the $x$-coordinate of each node in $J_k$. 

\begin{Lemma} \label{lem:xpositions}
Let $v_1=(A,B)$ be the transmission from node $A$ to node $B$, and let 
$v_2=(C,D)$ denote the transmission from node $C$ to node $D$.  Without loss of 
generality, suppose the second transmitter $C$ lies to the right of (or in the 
same position as) the first transmitter $A$ on the real number line, i.e. 
$x_{pos}(C) \ge x_{pos}(A)$.  Then, in the conflict graph constructed from 
conditions (a)-(e) of Construction~\ref{construction:KoseMedard} above, 
\\(1) transmissions $v_1$ and $v_2$ are adjacent vertices if and only if 
$x_{pos}(C) \in [x_{pos}(A), 2x_{pos}(B) - x_{pos}(A)]$.  
\\(2) transmissions $v_1$ and $v_2$ are nonadjacent in the conflict graph if 
and only if $x_{pos}(C) > x_{pos}(B) + s$, where $s$ is the distance between 
nodes $A$ and $B$.  
\end{Lemma}

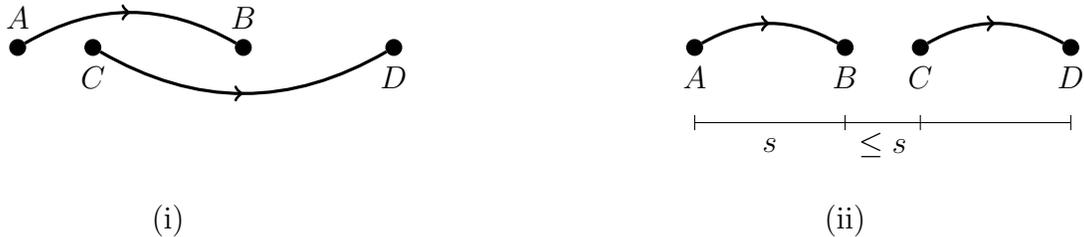
\begin{figure*}
\begin{center}
\begin{tikzpicture}
              
\vertex[fill] (A1) at (0,1) [label=above:$A$] {};
\vertex[fill] (B1) at (3,1) [label=above:$B$] {};
\vertex[fill] (C1) at (1,1) [label=below:$C$] {};
\vertex[fill] (D1) at (5,1) [label=below:$D$] {};

\vertex[fill] (A2) at (9,1) [label=below:$A$] {};
\vertex[fill] (B2) at (11,1) [label=below:$B$] {};
\vertex[fill] (C2) at (12,1) [label=below:$C$] {};
\vertex[fill] (D2) at (14,1) [label=below:$D$] {};

\draw[very thick, ->-=.5] (A1) to [bend left] (B1);
\draw[very thick, ->-=.5] (C1) to [bend right] (D1);

\draw[very thick, ->-=.5] (A2) to [bend left] (B2);
\draw[very thick, ->-=.5] (C2) to [bend left] (D2);

\node at (2, -1.3) {(i)};
\node at (11, -1.3) {(ii)};

\draw (9, -0) -- (14, 0);
\draw (9, -0-0.1) -- (9, -0+0.1);
\draw (11, -0-0.1) -- (11, -0+0.1);
\draw (12, -0-0.1) -- (12, -0+0.1);
\draw (14, -0-0.1) -- (14, -0+0.1);

\node at (10, -0.3) {$s$};
\node at (11.5, -0.3) {$\le s$};

% 
% \node at (10, 0) {$s$};
% \node at (11.5, 0) {$\le s$};

\end{tikzpicture}
\caption{Conditions under which receiver $B$ experiences interference due to 
transmitter $C$.}
\label{fig:proof:interference}
\end{center}
\end{figure*}

\noindent \emph{Proof}: Construction~\ref{construction:KoseMedard} is such that two transmissions $(A,B)$ and $(C,D)$ will be nonadjacent in the conflict graph 
only if receiver $B$ is closer to transmitter $A$ than to transmitter $C$.  
This condition is violated precisely when either (i) the intervals 
$[x_{pos}(A), x_{pos}(B)]$ and intervals $[x_{pos}(C), x_{pos}(D)]$ overlap, or 
(ii) if these two intervals are disjoint and the distance from $B$ to $C$ is 
at most the distance $s$ from $B$ to $A$; see 
Figure~\ref{fig:proof:interference}.  In case (i), there is interference at receiver $B$ because it is closer to transmitter $C$ than to transmitter $A$.  Case (ii) occurs exactly when $x_{pos}(C) 
\le x_{pos}(B) + s$.  This proves (1), and (2) follows immediately.
\qed

\begin{Theorem} \label{thm:countereg:KoseMedard}
Let $r_T$ denote the transmission radius of each wireless node.  Consider a 
line network with nodes $A_1, A_2, \ldots, A_n$ positioned on the $x$-axis in 
such a manner that each node $A_i$ is able to transmit to at most $2$ nodes to 
its right, i.e. $x_{pos}(A_{i+3}) - x_{pos}(A_i) > r_T$ for all 
$i=1,2,\ldots,n-3$.  Let $G_c$ denote the conflict graph of a set of valid 
transmissions in this network.  Then, there exists a line network for which the conflict graph $G_c$ contains a claw. 
\end{Theorem}

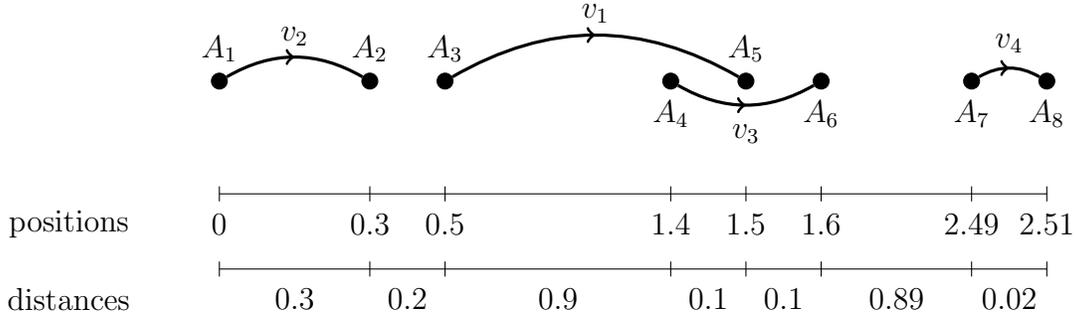
\begin{figure*}
\begin{center}
\begin{tikzpicture}              s
\vertex[fill] (A1) at (0,1) [label=above:$A_1$] {};
\vertex[fill] (A2) at (2,1) [label=above:$A_2$] {};
\vertex[fill] (A3) at (3,1) [label=above:$A_3$] {};
\vertex[fill] (A4) at (6,1) [label=below:$A_4$] {};
\vertex[fill] (A5) at (7,1) [label=above:$A_5$] {};
\vertex[fill] (A6) at (8,1) [label=below:$A_6$] {};
\vertex[fill] (A7) at (10,1) [label=below:$A_7$] {};
\vertex[fill] (A8) at (11,1) [label=below:$A_8$] {};

\draw[very thick, ->-=.5] (A1) to [bend left] (A2);
\draw[very thick, ->-=.5] (A3) to [bend left] (A5);
\draw[very thick, ->-=.5] (A4) to [bend right] (A6);
\draw[very thick, ->-=.5] (A7) to [bend left] (A8);

\node at (1, 1.6) {$v_2$};
\node at (5, 1.9) {$v_1$};
\node at (7, 0.3) {$v_3$};
\node at (10.5, 1.5) {$v_4$};
%===========================
\node at (-2, -0.9) {positions};

\draw (0, -0.5) -- (11, -0.5);
\draw (0, -0.5-0.1) -- (0, -0.5+0.1);
\draw (2, -0.5-0.1) -- (2, -0.5+0.1);
\draw (3, -0.5-0.1) -- (3, -0.5+0.1);
\draw (6, -0.5-0.1) -- (6, -0.5+0.1);
\draw (7, -0.5-0.1) -- (7, -0.5+0.1);
\draw (8, -0.5-0.1) -- (8, -0.5+0.1);
\draw (10, -0.5-0.1) -- (10, -0.5+0.1);
\draw (11, -0.5-0.1) -- (11, -0.5+0.1);

\node at (0, -0.9) {$0$};
\node at (2, -0.9) {$0.3$};
\node at (3, -0.9) {$0.5$};
\node at (6, -0.9) {$1.4$};
\node at (7, -0.9) {$1.5$};
\node at (8, -0.9) {$1.6$};
\node at (10, -0.9) {$2.49$};
\node at (11, -0.9) {$2.51$};

%===========================
\node at (-2, -1.9) {distances};

\draw (0, -1.5) -- (11, -1.5);
\draw (0, -1.5-0.1) -- (0, -1.5+0.1);
\draw (2, -1.5-0.1) -- (2, -1.5+0.1);
\draw (3, -1.5-0.1) -- (3, -1.5+0.1);
\draw (6, -1.5-0.1) -- (6, -1.5+0.1);
\draw (7, -1.5-0.1) -- (7, -1.5+0.1);
\draw (8, -1.5-0.1) -- (8, -1.5+0.1);
\draw (10, -1.5-0.1) -- (10, -1.5+0.1);
\draw (11, -1.5-0.1) -- (11, -1.5+0.1);

\node at (1, -1.9) {$0.3$};
\node at (2.5, -1.9) {$0.2$};s
\node at (4.5, -1.9) {$0.9$};
\node at (6.5, -1.9) {$0.1$};
\node at (7.5, -1.9) {$0.1$};
\node at (9, -1.9) {$0.89$};
\node at (10.5, -1.9) {$0.02$};

\end{tikzpicture}
\caption{A line network whose conflict graph contains a claw. The $x$-coordinates of nodes and distances between adjacent nodes are also shown.}
\label{fig:proof:counterexample:claw}
\end{center}
\end{figure*}

\noindent \emph{Proof:} Consider the line network shown in 
Figure~\ref{fig:proof:counterexample:claw}, consisting of $8$ nodes $A_1, 
\ldots, A_8$, positioned at locations $0, 0.3, 0.5, 1.4$, $1.5, 1.6, 2.49$ and 
$2.51$, respectively. Let the transmission radius of each node be $r_T=1$.  The 
nodes in these positions satisfy the condition that each node can communicate 
to at most two nodes to its right.  It can be verified that the $4$ 
transmissions $v_1=(A_3,A_5)$, $v_2=(A_1,A_2)$, $v_3=(A_4,A_6)$ and 
$v_4=(A_7,A_8)$ are valid (in the sense that each receiver is within communication radius of its transmitter), and that they form a claw in the conflict graph with 
center vertex $v_1$.  %xyz can add a figure for this
\qed

The problem of bounding the induced star number of the conflict graph becomes easier if it can be assumed that the set of valid transmissions consists of distinct unicast transmissions.  The next two results establish that this assumption can be made without loss of generality because the induced star number remains the same if multicast transmissions are replaced with corresponding unicast transmissions. 

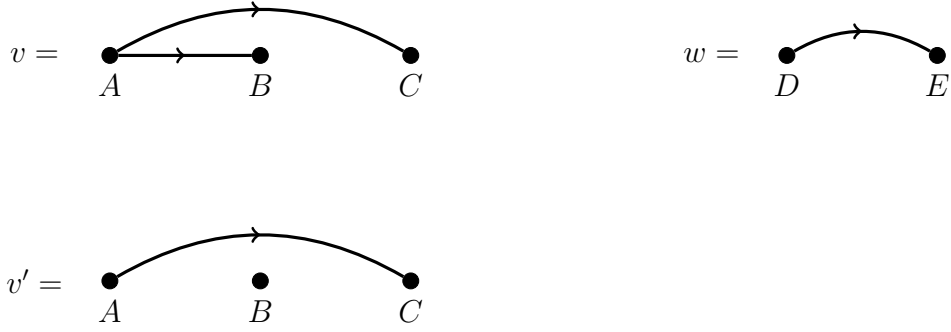
\begin{figure*}
\begin{center}
\begin{tikzpicture}
              
\vertex[fill] (A1) at (0,1) [label=below:$A$] {};
\vertex[fill] (B1) at (2,1) [label=below:$B$] {};
\vertex[fill] (C1) at (4,1) [label=below:$C$] {};

\vertex[fill] (A2) at (0,-2) [label=below:$A$] {};
\vertex[fill] (B2) at (2,-2) [label=below:$B$] {};
\vertex[fill] (C2) at (4,-2) [label=below:$C$] {};

\draw[very thick, ->-=.5] (A1) to [bend left] (C1);
\draw[very thick, ->-=.5] (A1) to (B1);

\draw[very thick, ->-=.5] (A2) to [bend left] (C2);

\vertex[fill] (D1) at (9,1) [label=below:$D$] {};
\vertex[fill] (E1) at (11,1) [label=below:$E$] {};

\draw[very thick, ->-=.5] (D1) to [bend left] (E1);

\node at (-1, 1) {$v = $};
\node at (-1, -2) {$v'=$};
\node at (8, 1) {$w = $};

\end{tikzpicture}
\caption{Transmissions $v$ and $w$ interfere with each other iff transmissions $v'$ and $w$ interfere with each other.}
\label{fig:proof:unicast:multicast}
\end{center}
\end{figure*}

\begin{Lemma} \label{lem:multicast:unicast}
Consider a line network having a valid multicast transmission $v=(A, \{B,C\})$ and a valid unicast transmission $w=(D,E)$, where the $x$-coordinate of node $B$ is less than the $x$-coordinate of node $C$; see Figure~\ref{fig:proof:unicast:multicast}.  Define a new unicast transmission $v'=(A,C)$.  Suppose Construction~\ref{construction:KoseMedard} is used to determine which pairs of transmissions interfere with each other.  Then, unicast transmission $w$ and multicast transmission $v$ interfere with each other if and only if unicast transmission $w$ and unicast transmission $v'$ interfere with each other.  
\end{Lemma}

\noindent \emph{Proof:}
Consider the intervals $\alpha = [x_{pos}(A), x_{pos}(C)]$ and $\delta = [x_{pos}(D), x_{pos}(E)]$.  As one slides $\delta$ from left to right on the $x$-axis, a few cases arise. 

First, suppose $\delta$ is to the left of and disjoint from $\alpha$. Then,  transmissions $w$ and $v$ do not interfere with each other if and only if receiver $E$ is closer to transmitter $D$ than to transmitter $A$, which is the case if and only if transmissions $w$ and $v'$ do not interfere with each other.  Observe that the fact that node $B$ is not a receiver in transmission $v'$ does not play a role in determining whether the two transmissions $w$ and $v$ (or $w$ and $v'$) interfere with each other because any interference between the two transmissions occurs at receivers $C$ or $E$ and is due to transmitters $A$ or $D$.  

Next, suppose $x_{pos}(E) = x_{pos}(A)$. Then by rule (b) of Construction~\ref{construction:KoseMedard}, transmissions $w$ and $v$ interfere with each other, as do transmissions $w$ and $v'$.  

If $x_{pos}(E) \in \alpha$ and $E \ne B$, then there exist two distinct nodes, namely $E$ and $B$, between transmitter $A$ and receiver $C$, contradicting the fact that a node can't transmit to more than two nodes to its right.  

Now suppose $x_{pos}(E) \in \alpha$ and $E=B$. Because node $A$ can't transmit to more than two nodes to its right, one has $x_{pos}(D) \le x_{pos}(A)$, and so receiver $E$ will experience interference due to transmitter $A$.  

Finally, if $x_{pos}(E) \ge x_{pos}(C)$, then any interference between transmissions $w$ and $v$ is due to interference at receiver $C$, in which case $w$ and $v'$ also interfere with each other.
\qed

Let $G, H$ be simple, undirected graphs, and let $v \in V(G)$. Let $G[v \leftarrow H]$ be the graph obtained by taking the disjoint union of $G-v$ and $H$, and joining each vertex in $H$ to each vertex in $G_1(v)$, where $G_1(v)$ denotes the neighbors in $G$ of vertex $v$. We say $G[v \leftarrow H]$ is the graph obtained from $G$ by \emph{replacing} vertex $v$ by $H$. 

\begin{Lemma} \label{lem:replacing:preserves:sigma}
 Let $G$ be a simple, undirected graph, and let $G[v \leftarrow K_r]$ denote the graph obtained from $G$ by replacing vertex $v$ of $G$ by the complete graph $K_r$ $(r \ge 1)$.  Then $\sigma(G[v \leftarrow K_r])=\sigma(G)$. 
\end{Lemma}

\noindent \emph{Proof:}
Recall that $\sigma(G)$ is defined as the number of leaf vertices in a maximum sized induced star in $G$.  If $v$ is the center vertex of a maximum sized induced star of $G$, then replacing $v$ by $K_r$ will replace the star with $r$ stars of the same size. Suppose $v$ is a leaf vertex of a maximum sized induced star of $G$. Since at most one vertex from a clique can belong to an independent set, and because the neighbors of $K_r$ in $G[v \leftarrow K_r]$ are the same as the neighbors of $v$ in $G$, we have $\sigma(G[v \leftarrow K_r])=\sigma(G)$. 
\qed

\begin{Remark} \upshape \label{rem:multicast:unicast}
Lemma~\ref{lem:multicast:unicast} says that in the conflict graph of a line network, vertices $v=(A, \{B,C\})$ and $w=(D,E)$ interfere with each other if and only if vertices $v'=(A, C)$ and $w=(D,E)$ interfere with each other.  A similar proof can be given to show that vertices $v=(A,\{B,C\})$ and $w=(D,\{E,F\})$ interfere with each other if and only if vertices $v'=(A,C)$ and $w'=(D,F)$ interfere with each other.   Thus, when constructing the conflict graph of a line network, it can be assumed that the given set $\{(i_k, J_k): k \in K\}$ of valid transmissions is such that each transmission is a unicast transmission, i.e. that $|J_k|=1$, for all $k \in K$.  If the original set of valid transmissions contained both a multicast transmission $(A, \{B,C\})$ and the corresponding unicast transmission $(A,C)$, then the new set of valid transmissions would contain two copies of the unicast transmission $(A,C)$.  Lemma~\ref{lem:multicast:unicast} implies that the conflict graph constructed for this new set of (unicast) transmissions is isomorphic to the conflict graph constructed for the original set.  Furthermore, by Lemma~\ref{lem:replacing:preserves:sigma}, replacing a single vertex in a graph with a complete graph on two vertices preserves the induced star number of the graph.  If there are two vertices in the conflict graph corresponding to the same unicast transmission, one of these vertices can be removed without affecting the induced star number. Hence, as far as results (or bounds) on the induced star number of the conflict graph of a line network are concerned, it can be assumed without loss of generality that the given set of valid transmissions consists only of distinct unicast transmissions. \qed
\end{Remark}

\begin{Theorem} \label{thm:line:networks:sigma:le:3}
Consider a line network whose nodes are $A_1, A_2, \ldots,A_n$, in order from left to right.  Suppose each node 
can transmit to at most two nodes to its right and all transmissions are in the 
direction from $A_1$ to $A_n$.  Then, under the protocol interference model (Construction~\ref{construction:KoseMedard}), the induced star number of the conflict graph of this line network 
is at most $3$.  Further, this bound is 
best possible.
\end{Theorem}

\noindent \emph{Proof:}  Without loss of generality, assume the transmission 
radius is $1$.  By way of contradiction, suppose the star $K_{1,4}$ is an 
induced subgraph of the conflict graph $G_c$, with center vertex $v_1$ and leaf 
vertices $v_2,v_3, v_4$ and $v_5$.  By Lemma~\ref{lem:multicast:unicast} and Remark~\ref{rem:multicast:unicast}, it can be assumed without loss of generality that the given set of valid transmissions consists of distinct unicast transmissions. Hence, we may assume that the vertices $v_i$ correspond to distinct unicast transmissions. 

\begin{figure*}
\begin{center}
\begin{tikzpicture}              
\vertex[fill] (A2) at (0,1) [label=below:$A_2$] {};
\vertex[fill] (B2) at (1,1) [label=below:$B_2$] {};
\vertex[fill] (A3) at (3,1) [label=below:$A_3$] {};
\vertex[fill] (B3) at (4,1) [label=below:$B_3$] {};
\vertex[fill] (A4) at (6,1) [label=below:$A_4$] {};
\vertex[fill] (B4) at (7,1) [label=below:$B_4$] {};
\vertex[fill] (A5) at (9,1) [label=below:$A_5$] {};
\vertex[fill] (B5) at (10,1) [label=below:$B_5$] {};

\draw[very thick, ->-=.5] (A2) to [bend left] (B2);
\draw[very thick, ->-=.5] (A3) to [bend left] (B3);
\draw[very thick, ->-=.5] (A4) to [bend left] (B4);
\draw[very thick, ->-=.5] (A5) to [bend left] (B5);

\draw (0, -0.5) -- (10, -0.5);
\draw (0, -0.5-0.1) -- (0, -0.5+0.1);
\draw (1, -0.5-0.1) -- (1, -0.5+0.1);
\draw (3, -0.5-0.1) -- (3, -0.5+0.1);
\draw (4, -0.5-0.1) -- (4, -0.5+0.1);
\draw (6, -0.5-0.1) -- (6, -0.5+0.1);
\draw (7, -0.5-0.1) -- (7, -0.5+0.1);
\draw (9, -0.5-0.1) -- (9, -0.5+0.1);
\draw (10, -0.5-0.1) -- (10, -0.5+0.1);

\node at (0, -1) {$a_2$};
\node at (1, -1) {$b_2$};
\node at (3, -1) {$a_3$};
\node at (4, -1) {$b_3$};
\node at (6, -1) {$a_4$};
\node at (7, -1) {$b_4$};
\node at (9, -1) {$a_5$};
\node at (10, -1) {$b_5$};

\node at (0.5, 1.5) {$v_2$};
\node at (3.5, 1.5) {$v_3$};
\node at (6.5, 1.5) {$v_4$};
\node at (9.5, 1.5) {$v_5$};

\end{tikzpicture}
\caption{Four mutually noninterfering transmissions.}
%xyz In figure above, put braces above the symbols for s and \le s
\label{fig1:proof:sigma:le:3}
\end{center}
\end{figure*}
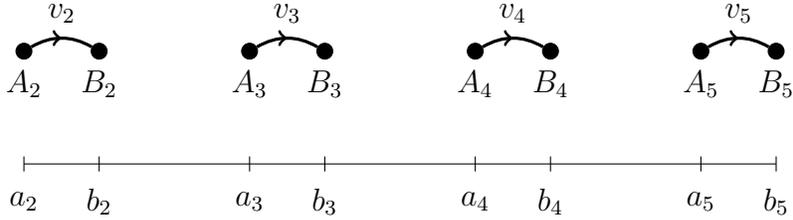

Suppose each transmission $v_i$ is a unicast transmission 
from a node $A_i$ at position $a_i$ to a 
node $B_i$ at position $b_i$.   Recall from the proof of 
Lemma~\ref{lem:xpositions} that if transmissions $v_i$ and $v_j$ are 
nonadjacent in $G_c$, then the closed intervals $[a_i,b_i]$ and $[a_j,b_j]$ are 
disjoint.  Since $\{v_2,v_3,v_4,v_5\}$ is an independent set in $G_c$, the closed 
intervals $[a_i,b_i]$, for $i=2,3,4,5$, are disjoint. Without loss of 
generality, assume the closed intervals are in order $v_2,v_3,v_4,v_5$ from left 
to right, as shown in Figure~\ref{fig1:proof:sigma:le:3}.  

Now consider a few cases, depending on the location $a_1$ of the transmitter 
$A_1$ of the transmission $v_1=(A_1,B_1)$.  Let $s_i := b_i -a_i$ denote the 
distance between the transmitter and receiver for transmission $v_i$, $i=1,\ldots,5$.  Since 
$v_2 v_3 \notin E(G_c)$, by Lemma~\ref{lem:xpositions} one obtains $a_3-b_2 > s_2$. We 
claim $a_1 < a_3$.  If $a_1 \ge a_3$, then $a_1-b_2 \ge a_3 - b_2 > s_2$, 
whence $v_1 v_2 \notin E(G_c)$ by Lemma~\ref{lem:xpositions}, a contradiction.  
Thus, $a_1 < a_3$.   Also,  if $b_1 > b_3$, then the four nodes $A_1, A_3, B_3, 
B_1$ are positioned in that order from left to right, which implies that there 
exists a node, namely $A_1$, which is able to communicate with up to $3$ nodes 
to its right, a contradiction.  Hence, $b_1 \le b_3$.  But then $a_5 - b_3 > 1$ 
and $b_1 \le b_3$ imply $a_5 - b_1 > 1$, whence $v_1 v_5 \notin E(G_c)$, a 
contradiction.  

It has been proved that the conflict graph $G_c$ does not contain a $K_{1,4}$ as 
an induced subgraph.  Hence, $\sigma(G_c) \le 3$.  This bound is best possible 
because, by Theorem~\ref{thm:countereg:KoseMedard} there exists a line network for which the conflict graph $G_c$ contains a claw. 
\qed

%====================================
\section{Concluding Remarks} \label{sec:concluding:remarks}

A general open problem is: what are the fundamental limits to the performance that is achievable with some given amount of resources?  More specifically, if each node in the wireless network has information only up to $d$ hops away, then what are the limits to performance? This problem was studied under the primary interference model, and sufficient conditions for admission control were obtained.  A distance-$d$ distributed algorithm was proposed.  It was shown that in the worst case, a distance-$d$ distributed algorithm can overestimate the amount of resources required to satisfy the given QoS bandwidth requirements by a factor of up to $(2d+3)/(2d+2)$.  This bound on performance is independent of the structure of the network graph.  This resolves an open problem posed in \cite{Ganesan:WN:2014}.  It was also shown that the complexity of the distributed algorithm executed at each node in the network to compute its local estimate is $O(\Delta^{5d})$, where $\Delta$ denotes the maximum degree of a vertex in the network graph, and so there is a tradeoff between performance and complexity. 

It was shown that for line networks, under the protocol interference model, the row constraints are a factor of at most $3$ away from optimal and that this bound is best possible. A line network was given for which the conflict graph contains a claw.  This implies that the polynomial time scheduling algorithms in the literature devised  for claw-free graphs are not directly applicable to solving the scheduling problem.  

These results can be extended in several directions.  First, in the  distance-$d$ distributed proposed in this work, it was assumed that wireless interference was modeled as primary interference.   This can be extended to conflict graphs constructed from other interference models.  Second, these results can be extended to more general interference models such as hypergraphs. For instance, when the interference is such that any two of some three wireless links can be simultaneously active, a hyperedge consisting of the three links captures the minimal forbidden set of links.  Third, the single-hop case was considered in the present work, and one can extend these results to the setting of multi-hop networks.  Fourth, the results on line networks under the protocol interference model can be extended to other topologies.  Finally, much work has been done on computing maximum weight independent sets in claw-free graphs and its special case of line graphs; the induced star number of these graphs is at most $2$. Designing efficient, distributed scheduling algorithms for graphs having bounded induced star number is an open direction.

%====================================
\section*{Acknowledgements} \label{sec:ack}

Thanks are due to the anonymous reviewers for helpful comments and suggestions.

 {
\bibliographystyle{plain}
\bibliography{refs_ag.bib}

}
\end{document}